\documentclass[showpacs,prd,twocolumn,aps]{revtex4}
\usepackage{graphicx}
\usepackage{psfrag}
\usepackage{bm}
\usepackage{epsfig}
\usepackage{amsmath}
\usepackage{amssymb}
\usepackage{psfrag}
\makeatother
\begin{document}

\title{{}``Cosmological'' quasiparticle production in harmonically trapped
superfluid gases}

\author{Petr O. Fedichev$^{1,2}$ and Uwe R. Fischer$^{1}$}

\affiliation{$^{1}$Leopold-Franzens-Universit\"{a}t Innsbruck, Institut f\"{u}r
Theoretische Physik, Technikerstrasse 25, A-6020 Innsbruck, Austria
\\
 $^{2}$Russian Research Center Kurchatov Institute, Kurchatov Square,
123182 Moscow, Russia}

\begin{abstract}
We show that a variety 
of cosmologically motivated effective 
quasiparticle space-times can be produced
in harmonically trapped superfluid Bose and Fermi gases.
We study the analogue of cosmological particle production in these 
effective space-times, induced by trapping potentials and coupling 
constants possessing an arbitrary time dependence. 
The WKB probabilities for phonon creation from the 
superfluid vacuum are calculated, and an experimental procedure 
to detect quasiparticle production by
measuring density-density correlation functions is proposed.  
\end{abstract}
\maketitle
\section{Introduction} 
In a gravitational field with explicit time dependence in the metric,
particles and antiparticles can be simultaneously created by quantum
fluctuations from the vacuum. By the uncertainty principle, the time
scale of the system's evolution dictates the typical energy of the
particles produced 
\cite{BirrellDavies}.
The process of 
cosmological particle production, whose condensed matter analogue
we shall consider here, is potentially relevant in the expanding early universe, 
in which phonons experience an acoustic geometry;
as a consequence, the expansion of the universe could generate density
waves growing into galaxies \cite{Sachs}.

Attention has of late focused on condensed matter analogs
of the curved space-times familiar from gravity, primarily due to
their conceptual simplicity and realizability in the laboratory \cite{GrishaPhysicsReports,Artifical,Wisdom,BoseCondensate,CSM,Leonhardt,PG,Schutzhold}.
Condensed matter systems lend themselves for an exploration of kinematical
properties of curved space-times and, in particular, provide a testbed
to study the effects of a well-defined and controlled {}``trans-Planckian''
physics, i.e. atomic many-body physics on a microscopic scale, 
on low-energy quantum effects
like Hawking radiation \cite{hawking} and cosmological particle
production. In the present paper, we investigate  
quantum fields propagating on effective curved space-times backgrounds,  
for the case of harmonically trapped, dilute superfluid 
gases with possibly time varying particle interactions.
For a perfect, irrotational liquid, described by Euler and continuity
equations, it was recognized by Unruh \cite{unruh} that the action of 
fluctuations of the velocity potential $\Phi$,
around a spatially inhomogeneous and time dependent background, can be
identified with the action of a minimally coupled scalar field according
to 
\begin{eqnarray}
S & = & \int dt d^3 x\frac{1}{2\kappa}\left[-\left(\frac{\partial}{\partial t}
{\Phi}-{\bm v}\cdot\nabla\Phi\right)^{2}
+c^{2}(\nabla\Phi)^{2}\right]\nonumber \\
 & \equiv & \frac{1}{2}\int dtd^3 x
\sqrt{-{\textsf{g}}}{\textsf{g}}^{\mu\nu}\partial_{\mu}\Phi\partial_{\nu}\Phi\,.\label{action}
\end{eqnarray} 
Here, $\bm v$ is the background velocity, $1/\kappa$ the compressibility,
and $c$ the speed of sound of the liquid. 
We use the summation convention over
equal indices, unless indicated otherwise. 
The quantities ${\textsf g}^{\mu\nu}$ 
are the {\em contravariant} components of the effective 
metric tensor related to its covariant components by 
$\textsf{g}^{\beta\nu}\textsf{g}_{\nu \alpha}= \delta^{\beta}{}_\alpha$, 
and ${\textsf g} \equiv {\rm det}\,{\textsf g}_{\mu\nu}$ 
is the determinant of the metric tensor. 
The action (\ref{action}) leads to the {}``relativistic'' scalar wave equation 
\begin{equation}
\square\,\Phi\equiv
\frac{1}{\sqrt{-\textsf{g}}}\partial_{\mu}\left(\sqrt{-\textsf{g}}
\; \textsf{g}^{\mu\nu}\;\partial_{\nu}\Phi\right)=0. 
\label{E:dalembertian}
\end{equation}
In general, the effects of quantum fluctuations described by the quantum
version of Eq.\,(\ref{action}) are very small and can hardly be observed
because of finite temperature and dissipation effects. Therefore atomic 
superfluids, where both extremely small temperatures and 
dissipationless flows are possible, attract growing interest for an 
emerging research field of ``experimental cosmology.'' 

In the following, we study how various curved space-times 
can be implemented in harmonically 
trapped superfluid Bose and Fermi gases. 
As a concrete 
example, we show how de Sitter and Friedmann-Robertson-Walker (FRW) 
universes can be {}``re-created'' in superfluid gases.
We analyze the quasiparticle production probabilities, leading to 
a thermal spectrum in the WKB approximation, and discuss an experimental 
procedure to observe and characterize the excitations produced.

\section{Quasiparticle metric tensors in harmonically trapped superfluids}
\subsection{Superfluid action} 
The hydrodynamic, i.e. long-wavelength action of a trapped superfluid is generally 
given by 
\begin{equation}
S=\int dtd^3 x\left[\rho\frac{\partial}{\partial t} {\phi}
+\frac{\rho}{2}{(\nabla\phi)^{2}}+\epsilon(\rho)
+V_{\textrm{trap}}\rho\right],
\label{fundaction}
\end{equation}
where the external harmonic potential $V_{\textrm{trap}}({\bm x},t)
=\frac{1}{2}(\omega_{x}^{2}(t)x^{2}+\omega_{y}^{2}(t)y^{2}+\omega_{z}^{2}(t)z^{2})$
is characterized by the three frequencies $\omega_{i}$, 
$i=x,y,z$.  The trapping frequencies are assumed to be time dependent 
in an arbitrary manner 
(we can also conceive of making $\omega_{i}^{2}(t)$ effectively
negative by ``turning over'' the potential, see section \ref{deSitter}).
In the above action, the quantity $\rho$ plays the role of a 
response or {\em stiffness} coefficient to gradients of $\phi$, and equals
the total fluid density at absolute zero; the equation of state of the superfluid is given by the  
energy density functional $\epsilon=\epsilon(\rho)$.
(Note that we leave out an overall minus sign in the definition of the
action $S$.) We generally set $\hbar=m=1$. 
The action entails the existence of a velocity potential $\phi$, 
such that the vorticity is zero except on singular lines, and ensures the validity of 
Euler and continuity equations for the superfluid velocity ${\bm v}  =\nabla\phi$. 
The identification of $\phi$ with the phase of a complex 
``order parameter'' 
(i.e., the direction of a unit vector in the plane of some abstract space) 
leads to the quantization of circulation, because $\phi$ is then defined only modulo multiples 
of $2\pi$. Finally, the above action implies the conjugateness of phase and density 
quantum variables \cite{anderson}: 
\begin{equation}
[\rho ({\bm x}), \phi ({\bm x}')] = i\delta ({\bm x}-{\bm x}') \,. 
\label{comm} 
\end{equation} 
These properties, taken together, constitute the canonical definition of a superfluid
at $T=0$ \cite{khalatni,leggettBEC}. Therefore, Eq.\,(\ref{fundaction}) represents 
the universal action of a simple scalar superfluid at absolute zero
made up of elementary bosonic or fermionic atoms, independent of a 
particular microscopic model.

The simplest example for the equation of state 
is that of a weakly interacting Bose gas, 
with $\epsilon_{\textrm{B}}=\frac{1}{2}g\rho^{2}$,
where $g$ is the coupling constant, $g=4\pi a_s$, with $a_s$ the
$s$-wave scattering length, characterizing pair collisions of atoms.
The scattering length can be tuned
using external magnetic fields \cite{Feshbach}. Another example is
a two-component Fermi gas with attractive interactions between 
atoms of different hyperfine species \cite{OHara}. 
The ground state of such a gas is
superfluid (in the simplest version, it is the BCS state of a 
scalar superfluid with $s$-wave
pairing), and since the interactions are weak, the BCS gap is small
and the equation of state (to exponential accuracy) coincides
with that of a free Fermi gas:
$\epsilon_{\textrm{F}}=\left[{(3\pi^{2})^{2/3}}/{10}\right] \rho^{5/3}$. 
To consider all possible cases which have a power law 
density-dependence of the equation of state, in a generic 
way, we write 
\begin{equation}
\epsilon(\rho)=Ag^{\beta}\rho^{\gamma},
\end{equation} 
where $A$ is a numerical constant.
That is, $\beta=1,\gamma=2$ for a dilute Bose gas, 
and $\beta=0,\gamma=5/3$ for noninteracting two-component
fermions.

In our present context, an important quantity characterizing a superfluid is the
so-called {}``Planckian'' energy scale, $E_{\rm Pl},$ i.e. the 
frequency beyond which the spectrum of the excitations above the superfluid ground state 
ceases to be phononic and (pseudo-)Lorentz invariance is broken. 
For a weakly interacting Bose gas $E_{\rm Pl}\sim g\rho$,
of order the mean interparticle interaction. In a BCS superfluid,
$E_{\rm Pl}$ is determined by the BCS gap: 
$E_{\rm Pl}\sim\rho^{2/3}\exp[-1/(|g|\rho^{1/3})]$.
The complete analogy with a quantum field theory on a {\em fixed} 
curved space-time background given by (\ref{action}) only exists 
if all timescales $t_{0}$, describing the evolution of the superfluid, 
are much larger than the ``Planck time'': $t_{0}\gg 1/E_{\rm Pl}$.

\subsection{Scaling transformation for Bose and Fermi superfluids} 

Our approach in the following is based on the so-called scaling transformation 
\cite{Scaling,PitaRosch,ScalingG(t),Menotti}
to describe the expansion and contraction of the gas under time dependent
variations of the trapping frequencies. 
It is by now a well-established
fact that the hydrodynamic solution for density and velocity of motion
for such a system may be obtained from a given initial solution by
a scaling procedure both in the bosonic \cite{Scaling,PitaRosch,ScalingG(t)}
as well as in the fermionic case \cite{Menotti}. Defining the scaled
coordinate vector ${{\bm x}_{b}}={\bm e}_i x_i/b_i$, density and velocity
are given by the scaling  transformations  \cite{Scaling}:
\begin{eqnarray}
\rho({\bm x},t) & \Rightarrow & \frac{\tilde{\rho}({\bm x}_b)}{\mathcal{V}}\label{ScalDensity}\\
\phi({\bm x},t) & \Rightarrow & 
\frac{\dot{b}_{i}}{2b_{i}}x_{i}^{2}
+\tilde{\phi}({\bm x}_{b},t)\,.\label{ScalVelocity}
\end{eqnarray}
 The (dimensionless) scaling volume ${\mathcal{V}}=\prod_{i}b_{i}$
in the density (\ref{ScalDensity}) is dictated by particle conservation.


Introducing a new ``scaling time'' variable by 
\begin{equation}
\frac{d\tau_s}{dt}=\frac{(g/g(0))^{\beta}}{{\mathcal{V}}^{\gamma-1}} , 
\label{deftau}
\end{equation}
 we rewrite the action (\ref{fundaction}) in the form 
\begin{eqnarray}
S & = & \int d\tau_s d^3 x_{b}\left[\tilde{\rho}\frac{\partial}{\partial\tau_s}\tilde{\phi}
+\frac{\tilde{\rho}}{2}F_{i}(\tau_s){(\nabla_{bi}\tilde{\phi})^{2}}\right.\nonumber \\
 &  & \hspace*{5em}\left.+\tilde{\epsilon}(\tilde{\rho})+V_{\textrm{trap}}({\bm x}_{b},0)\tilde{\rho}\right],
\end{eqnarray}
 where the $\tau_s$ dependent {\em scaling factors}  are  
\begin{equation}
F_{i}(\tau_s)=\frac{{\mathcal{V}}^{\gamma-1}}{b_{i}^{2}(g/g(0))^{\beta}}
= \frac{1}{b_{i}^{2}}\frac{dt}{d\tau_s},
\label{defF}
\end{equation}
 and $\tilde{\epsilon}(\tilde{\rho})=\epsilon(\tilde{\rho})_{|g=g(0)}$; 
$\nabla_{bi}\equiv \partial/\partial x_{bi}$. 
The rescaled density $\tilde{\rho}$ has no explicit $\tau_s$ dependence
(whereas it has explicit dependence on the lab time $t$), 
and coincides with the equilibrium condensate density profile in the scaling
coordinate ${\bm x}_b$.
Where any confusion might arise, we will generally designate 
scaling variables with a tilde to clearly distinguish them from
 lab frame variables.

For the relation (\ref{defF}) between the scaling factors and ${b_i},g$
to hold true, we must impose the following equations of motion 
for the scaling parameters $b_{i}$: 
\begin{equation}
\ddot{b}_{i}+\omega_{i}^{2}(t)b_{i}=\frac{(g/g(0))^{\beta}\omega_{0}^{2}}{{\mathcal{V}}^{\gamma-1}b_{i}}.
\label{genbEq}
\end{equation}
They need to be solved with the initial conditions $b_{i}=1$
and $\dot{b}_i=0$. Note that here no summation convention is used
in the second term on the LHS. For a sufficiently large cloud, the 
stationary background solution can be found from the 
Thomas-Fermi density profile. 
It is given by using that $\mu=-d\tilde\phi/d\tau_s$ equals the initial chemical
potential, and 
\begin{equation}
\frac{d\tilde{\epsilon}}{d\tilde{\rho}}=\mu-V_{\rm trap}({\bm x}_{b},0)\,.\end{equation}
The part of the action quadratic in the fluctuations is obtained to be 
\begin{eqnarray}
S^{(2)} & = & \int d\tau_s d^3 x_{b}\left[\delta\tilde{\rho}\frac{\partial}{\partial\tau_s}\delta\tilde{\phi}
+\frac{\tilde{\rho}}{2}F_{i}{(\nabla_{bi}\delta\tilde{\phi})^{2}}
+\frac{1}{2}\tilde{\kappa}\delta\tilde{\rho}^{2}\right],\nonumber \\ 
\end{eqnarray}
where $\delta\tilde{\rho}=\tilde{\rho}-\tilde{\rho}_{0}$ and 
$\delta\tilde{\phi}=\tilde\phi+\mu\tau_s$. 
The rescaled bulk compressional modulus (inverse compressibility) 
$\tilde{\kappa}={d^{2}\tilde{\epsilon}}/{d^{2}\tilde{\rho}}$
does not depend on the time $\tau_s$, and is identical to $g(0)$ in the
bosonic case. After integrating out the density
fluctuations, we obtain the effective action for the rescaled phase variable
\begin{equation}
\bar{S}^{(2)}=\int d\tau_s d^3{x_b}\frac{1}{2\tilde{\kappa}}
\left[-\left(\frac{\partial}{\partial\tau_s}\delta\tilde{\phi}\right)^{2}
+ 
\tilde c^2 F_{i}(\nabla_{bi}
\delta\tilde{\phi})^{2}\right],\label{kappaaction}
\end{equation}
where the squared scaling speed of sound 
${\tilde c}^2=\tilde \kappa \tilde \rho ({\bm x}_b)$. 
Using the identification with a minimally coupled scalar field, 
analogous to the one performed in the second line of 
Eq.\,(\ref{action}), the line 
element in the scaling variables reads 
\begin{equation}
ds^{2}=\frac{\tilde c}{\tilde{\kappa}}
\sqrt{F_{x}F_{y}F_{z}} \left[-{\tilde c}^{2}d\tau_s^{2}
+F_{i}^{-1}dx_{bi}^{2}\right].\label{blineelement}
\end{equation}
The line element 
takes a particularly simple form for an isotropic superfluid Fermi gas, 
where $\gamma=5/3$ 
and thus all $F_{i}=1$, 
leading to $d\tau_s/dt=b^{-2}$. We note that even for this simple case, 
the metric defined by (\ref{blineelement}) is not trivial, since $\tau_s$ and
$t$ are different, and both $\tilde c$ and $\tilde \kappa$ depend on the radial 
scaling coordinate ${r}_{b}$. We will see below that in the case that $F_i=1$ 
the scaling transformation is exact, and that 
therefore no quasiparticle creation occurs 
in the {\em scaling variable basis}, i.e. there is no mixing of negative and
positive frequency parts in the time $\tau_s$ 
(quasiparticle creation can take place in the lab frame with time $t$, though,
and a lab detector will still see that quasiparticles are ``created'').

Identifying $\delta\tilde{\phi}$ with $\Phi$,
and going back from scaling coordinates to laboratory frame variables,
we recover the action (\ref{action}), with ${\bm v}=(\dot{b}_{i}/b_{i})r_{i}{\bm e}_{i}$,
and the line element, which is of Painlev\'{e}-Gullstrand type, reads
\cite{PGoriginal,Matt}
\begin{equation}
ds^{2}= 
\frac{c}{\kappa}
\left[-(c^{2}-{\bm v}^{2})dt^{2}-2v_{i}dx_{i}dt+dx_{i}^{2}\right],\label{lineelement}\end{equation}
where $c^{2}=\kappa\rho$ is the squared instantaneous speed of sound.
We now assume that space is spherically symmetric, i.e. that $\bm v$
has a radial component $v_{r}$ only, that $v_r/c = f(r)$ holds, 
and furthermore $c=c(t)$ is a function of time only.  
We first apply the transformation $c_0 d\tilde t = c(t) dt$,  
where $c_0$ is some constant (initial) sound speed, 
connecting the laboratory time $t$ to the time variable $\tilde t$.
This results in the line element  
$d s^2 
=(c/\kappa)[-c_0^2(1- f^2)d \tilde t^2 -2f c_0 dt dz +dr^2 +r^2d\Omega^2$.
We then employ a second transformation 
$c_0 d\tau= c_0 {d\tilde t} + f dr/(1-f^2)$ \cite{note}, 
to bring the metric into the form
\begin{equation}
ds^{2}= 
\frac{c}{\kappa}
\left[-\left(1-f^{2}(r)\right)c_0^2 d{\tau}^{2}
+\frac{1}{1-f^{2}(r)}dr^{2}
+r^{2}d\Omega^{2}\right],\label{diagonallineelement}
\end{equation}
The metric in the above form 
facilitates comparison with metric tensors in spherically
symmetric space-times written in their standard form. E.g., if 
$f^2(r) = 2M/r$ is chosen, this line element
is conformally equivalent to the Schwarzschild metric, the asymptotically
flat vacuum solution of the Einstein equations around a spherically symmetric 
body with total mass $M$ \cite{Weinberg}.

\section{Creating De Sitter and Friedmann-Robertson-Walker 
universes} 
\label{deSitter}

The equation of state for Bose superfluids contains the interatomic
interaction. Therefore, by varying this interaction, possibly together
with the trapping frequencies, 
expanding clouds of Bose atoms allow for the simulation 
of a large set of cosmological space-times. We begin by discussing
the so-called de Sitter universe, which is a solution of the vacuum
Einstein equations characterized by the line element \cite{deSitter,Weinberg}
\begin{equation}
ds^{2}=-c^{2}\left(1-\frac{\Lambda}{3}r^{2}\right)d{\tau}^{2}
+\left({1-\frac{\Lambda}{3}r^{2}}\right)^{-1}dr^{2}+r^{2}d\Omega^{2}\,,\label{deSitterlineelement}
\end{equation}
with $\Lambda$ being the cosmological constant 
$\equiv$ energy density of the vacuum. 
Up to the conformal factor $c/g$, the metric 
(\ref{diagonallineelement}) coincides with the de Sitter
metric (\ref{deSitterlineelement}), provided 
we require that $v_{r}^{2}/c^{2}=\Lambda r^{2}/3$
and that the speed of sound is a constant in space and time. 
The speed of sound in the center of the cloud 
is time independent if:
\begin{equation}
c^{2}= {\tilde c}^2={\textrm{const.}}\quad
\Longleftrightarrow\quad g(t)/g(0)=b^{3}(t)\,.\label{ccond}
\end{equation}
Close to the center of the condensate, $c$ is, in addition, practically 
spatially independent. 
Using Eq.\,(\ref{ScalVelocity}), we find that $\Lambda={\rm const}.$ provided
$b\propto\exp[\lambda t]$, with $\lambda=c\sqrt{\Lambda/3}$.
This exponential expansion of the cloud 
can (asymptotically) be achieved if we
turn over the potential, making it expel the particles rather than
trapping them: $\omega^{2}(t\rightarrow\infty)=-\lambda^{2}$. The
de Sitter horizon, where $v_{r}=c$, is stationary and situated at
$r_{h}=c(0)/\lambda$, which is well inside the expanding cloud provided 
$\lambda\gg\omega_{0}$, where $\omega_{0}$ is the initial
trap frequency. 

\begin{center}
\begin{figure}[hbt]
\psfrag{r=0}{\large $r=0$}
\psfrag{v_r=c_s}{\large $v_r=c_s$}
\psfrag{v_r}{\large $v_r$}
\psfrag{B}{\normalsize $\bm B$}
\psfrag{R}{\large $R$}
\psfrag{F=+1/2}{$F=+\frac12$}
 \psfrag{F=-1/2}{$F=-\frac12$}
\centerline{\epsfig{file=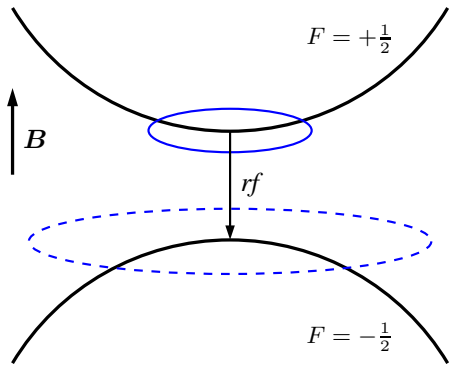,width=0.33\textwidth}}
\caption{\label{scheme} Exponential 
expansion of a two-component Bose-Einstein 
condensate with effective spin $F=1/2$, to create a de Sitter quasiparticle
universe close to its center. The trapping potential is inverted by applying 
a radio frequency ($r\!f$) pulse, which transfers the atoms to their untrapped
hyperfine state.}
\end{figure}
\end{center}

The experimental sequence leading to {}``condensate inflation''
is schematically 
depicted in Fig.\ref{scheme}. We assume that the experiment can
be done with one trapped (low-field seeking) and one untrapped (high-field
seeking) hyperfine component of the same atomic species. We start
from a sufficiently large Bose-Einstein-condensed cloud at small (effectively
zero) temperature with all atoms being in the trapped state. Then,
we transfer all the atoms to the untrapped state, by 
flipping the sign of the trapping potential. At the same time, we
ramp up the interaction strength, according to condition (\ref{ccond}),
using a suitable Feshbach resonance \cite{Feshbach}. As a result
of the simultaneous action of the inverted parabolic potential and
the increasing interaction energy, the gas cloud experiences a rapid
exponential expansion, representing the analogue of cosmological inflation.

In fact, Eq.\,(\ref{ccond}) defines a broad class
of Bose superfluid effective space-times. 
In the present case of isotropic expansion with $b_{i}=b$, we have $t=\tau_s$ 
and, in scaling variables, we obtain up to a conformal factor 
a Friedmann-Robertson-Walker metric:
\begin{equation}
ds^{2}=\frac{c}{\kappa b^{3}}
\left[-c^{2}dt^{2}+b^{2}d{r_{b}}^{2}+b^{2}{r_{b}}^{2}d\Omega^{2}\right].
\label{blineelementiso}
\end{equation}
According to the above form of the metric, the quantity $b$ plays
the role of the scaling parameter not only in our condensed matter
context, but can be interpreted equally well as the scale factor in
the expansion of the universe, with $H\equiv\dot{b}/b$ the Hubble
parameter. As demonstrated above, exponential growth of $b$, with
constant $H$, corresponds to exponential inflation \cite{inflation}.
The present setup also allows for the simulation of power law inflation
\cite{inflation,EMomtensor}, with $b(t)\propto t^{\delta}$. The
{}``Hubble parameter'' $H$ changes for all exponents $\gamma$
inversely proportional to time $t$, $H\propto1/t$, and the exponent
$\delta=1/2$ corresponds to a {}``radiation dominated'' universe,
while the exponent $\delta=2/3$ corresponds to a {}``matter dominated''
universe. An isotropically trapped expanding superfluid gas thus models
an isotropic expanding universe. Generically, we can model anisotropic
universes with (\ref{blineelement}), with scaling factors which are
different in different spatial directions.

\vspace*{-0.5em}
\vspace*{-2em}

While this experiment is feasible in principle, increasing the interaction
dramatically increases three-body losses as well, whose total rate 
scales like $g^4 \rho^2$. 
This complication can be avoided, by switching to effectively lower-dimensional
systems; e.g., a 1+1D analogue of a de Sitter universe can be
achieved for quasi-1D excitations in a linearly expanding elongated
Bose-condensate, without changing the interaction \cite{fedichev:hawkingPRL}.
Another possibility is to use superfluid Fermi gases. 
We have, in the isotropic case, $F=1$, $d\tau_s = dt/b^2$, and the metric may
be written in the form
\begin{equation}
ds^{2}=\frac{\tilde c}{\beta^2\tilde\kappa}
\left[-\tilde c^{2}dt^{2}+\beta^{2}d{r_{b}}^{2}+\beta^{2}{r_{b}}^{2}
d\Omega^{2}\right],
\label{Blineelementiso}
\end{equation}
where we defined a scale factor $\beta \equiv b^{2}$. 
Performing experiments in superfluid Fermi gases has the advantage
that three-body losses are strongly suppressed by Fermi statistics 
\cite{petrov:3bodyrecfermi}.

\section{Cosmological particle production analogue}

Now we turn to describe the evolution of quantum fluctuations, on top of the classical
(mean field) hydrodynamic solutions described above. 
The equation of motion for the phase fluctuations can be obtained
after variation of the action (\ref{kappaaction}), 
\begin{equation}
\frac{\partial^{2}}{\partial\tau_s^{2}}\delta\tilde{\phi}-\tilde{\kappa}\nabla_{bi}
\left(\frac{\tilde c^{2}}{\tilde{\kappa}}F_{i}(\tau_s) \nabla_{bi} 
\delta\tilde{\phi}\right)=0.
\end{equation}
Phrased in curved space-time language, 
the above equation is the minimally coupled massless scalar 
wave equation for $\delta\tilde\rho$, analogous to (\ref{E:dalembertian}), 
with the metric (\ref{blineelement}).

Consider for simplicity the isotropic case, 
\begin{equation} 
F_i \equiv F= \frac{b^{3\gamma - 5}}{(g/g(0))^\beta} .
\end{equation}
The solution for the full quantum field reads
\begin{eqnarray}
\delta\tilde{\phi} & = & \sum_{n}\sqrt{\frac{\tilde{\kappa}}{2\tilde{V}\epsilon_n}}\phi_{n}({\bm x}_{b})\left[{\hat{a}}_{n}\chi_{n}+{\hat{a}}_{n}^{\dagger}\chi_{n}^{*}\right]\,,\label{3DSol}
\end{eqnarray}
 where $\tilde{V}$ is the the initial Thomas-Fermi volume of the
cloud, the operators $\hat{a}_{n}$ ($\hat{a}_{n}^{\dagger}$) annihilate
(create) phonon excitations in the initial vacuum state, 
and the mode functions $\chi_{n}$ satisfy 
\begin{equation}
\frac{d^{2}}{d\tau_s^{2}}\chi_{n}+F(\tau_s)\epsilon_{n}^{2}\chi_{n}=0.\label{chin}\end{equation}
The initial conditions $\chi_{n}(\tau_s\rightarrow-\infty)=\exp[-i\epsilon_{n}\tau_s]$
are selected such that Eq.\,(\ref{3DSol}), at $t\rightarrow-\infty$, 
describes the phase fluctuations in a static trapped superfluid in
its ground state. In quantum field theory (QFT) language this
ensures that a laboratory frame detector does not detect quasiparticles
at $t\rightarrow-\infty$. Hereafter we define the ``scaling vacuum'' to be 
the quantum state annihilated by the operators $\hat{a}_{n}$, where our
choice of the initial conditions gurantees that the initial~superfluid
vacuum and the scaling vacuum coincide at $t\rightarrow-\infty$. 

The case when all $F_{i}\equiv 1$ is remarkably special: 
In this case Eq.\,(\ref{chin})
does not depend on the superfluid evolution, and thus the quantum
state of the excitations remains unchanged. As we have seen above,
this indeed happens in the case of an isotropic Fermi superfluid.
Another example is a 2D isotropic dilute Bose gas with constant particle 
interaction  \cite{PitaRosch}. 
In these cases the scaling transformation
is exact, both for the condensate and the excitations.
In the language of QFT this amounts to the fact that there is no particle
production in the scaling basis, since the scaling solution is 
constructed from eigenfunctions of the (exactly conserved) scaling transformation
operator $F_i$ (in other words the scaling vacuum is protected by 
an exact scaling invariance which forbids frequency mixing).
The fact that no excitations are produced
in the scaling basis does not mean that a lab
detector does not detect quasiparticles. The transformation from the
laboratory time $t$ to the scaling time $\tau_s$ time is nontrivial and 
thus the phase of the functions $\chi_n$ is a complicated function of the
laboratory time $t$. In other words,
the phase field given by Eq.\,(\ref{3DSol}),  
if coupled to a detector of the type considered in 
\cite{fedichev:hawkingPRL}, gives a non-zero response.
This indeterminacy of the vacuum state finds its counterpart in 
the Unruh-Davies effect in flat space-time \cite{unruh76,BirrellDavies}  
and its curved space-time generalization, the Gibbons-Hawking effect \cite{Gibbons}. 

The description of a quantum field state in terms of particles and antiparticles
is based upon the separation of positive and negative frequency parts. 
As we confine ourselves to a measurement
involving the laboratory time variable $t$, this distinction is only possible
if the asymptotic phase of $\chi_n$-functions is large and sufficiently
quickly increases as a function of $t$. Using a WKB approximation to
the solutions of Eq.\,(\ref{chin}), we find
\begin{equation}
\lim_{t\rightarrow\infty}\int^{t}\sqrt F\, \frac{d\tau_s}{dt}dt\neq0.
\label{TPS}
\end{equation}
 The latter condition can be also called a ``Trans-Planckian safety
condition'' (TP condition), since if fullfilled it implies that an experiment in a
lab frame probing an energy scale $E_{0}\ll E_{\rm Pl}$ does not require
information about solutions of Eq.\,(\ref{chin}) with $\epsilon_{n}\agt E_{\rm Pl}$.
For isotropic expansion of a 3D Bose gas, Eq.\,(\ref{TPS}) is equivalent
to divergence of $\int dtg^{1/2}/b^{5/2}$ and is quite restrictive:
For the FRW analogy discussed above avoiding the divergence implies,
according to (\ref{ccond}), that \textbf{$b$} should
not grow faster than linearly.
The TP condition (\ref{TPS}) is based upon the
WKB approximation condition for Eq.\,(\ref{chin}), leading to the 
requirement $F(\tau_s)\agt(\epsilon_{n}\tau_s)^{-2}$ for large $\tau_s$ (here $\agt$
means ``grows faster than''). Substituting the latter condition into
Eq.\,(\ref{TPS}) we find that the marginal WKB case corresponds to a 
logarithmically divergent integral in Eq.\,(\ref{TPS}). 
Thus the marginal TP case corresponds to the marginal WKB case and
vice versa. 

The equation (\ref{chin}) is formally equivalent to scattering of a non-relativistic
particle with energy $\epsilon_{n}$ by a potential $\epsilon_{n}^{2}(1-F(\tau_s))$.
The initial conditions correspond to a single particle per unit time
incident 
on the potential barrier. Time dependence
of the scaling factors leads to scattering of the particles from the
incoming wave and at $\tau_s\rightarrow\infty$ the WKB solution reads: 
\begin{equation}
\chi_{n}=\frac{1}{F^{{1}/{4}}}
\left(
\alpha_{n}e^{-i\epsilon_{n}\int{d\tau_s}\sqrt{F}}+\beta_{n}e^{i\epsilon_{n}\int{d\tau_s}\sqrt{F}}\right),
\label{chinEqs}
\end{equation}
 where $\alpha_{n}$ is the transmission and $\beta_{n}$ the forward
scattering amplitude. The coefficients $\alpha_{n}$ and $\beta_{n}$
are related via the particle flux conservation condition:
\begin{equation}
|\alpha_{n}|^{2}- |\beta_{n}|^{2}=1.\label{bognorm}\end{equation}
In QFT language the latter condition is the Bogoliubov
transformation normalization condition for a bosonic field. The number
of particles detected by a scaling time detector at rest
is measured by the absolute square $|\beta_{n}|^{2}$ (which is proportional
to the probability that the detector absorbs a quantum), and therefore
$N_{n}=|\beta_{n}|^{2}$ can be interpreted as the number of 
scaling basis quasiparticles created.

In the WKB approximation the amplitudes are connected in a simple
way:
\begin{equation}
\beta_{n}=\exp[-\epsilon_{n}/2T_{0}]\alpha_{n},
\end{equation}
where the inverse temperature is given by the integral 
\begin{equation}
\frac{1}{T_{0}}={\Im}\left[\int_{\cal C}\sqrt F \, d\tau_s\right], \label{defT0}
\end{equation}
and $\cal C$ is the contour in the complex $\tau_s$-plane enclosing the
closest to the real axis singular point of the function $F(\tau_s)$ \cite{ll}.
Together with Eq.\,(\ref{bognorm}), this gives 
\begin{equation}
N_n=|\beta_{n}|^{2}=\frac{1}{\exp[\epsilon_{n}/T_{0}]-1},
\end{equation}
i.e. adiabatic evolution of trapped gases leads to {}``cosmological''
quasiparticle creation with thermal occupation numbers in the scaling basis.
The temperature $T_{0}$ depends on the details of the scaling evolution 
(see the specific example in Eq.\,(\ref{temperature}) below).


Interestingly, the evolution of the scaling parameters $b_{i}$,
and therefore the nontrivial line element (\ref{blineelement}), can
be generated already in a non-expanding cloud with time-dependent interaction
$g(t)$. A similar experiment has been suggested
in \cite{BLVFRW}, where time dependent interactions were used
to simulate FRW cosmologies and quantum quasiparticle production. 
The difference to our approach is due to 
the fact that the authors of \cite{BLVFRW} consider a trap with very steep 
walls (effectively a hard-walled container), so that the density of the cloud does not change, 
and the superfluid velocity vanishes everywhere at all times. 
In our setup, we are able to induce cosmological quasiparticle production 
in a harmonically trapped gas, by changing 
simultaneously the harmonic trapping and the interaction. 
The simplest case we can consider is to leave all $b_i=1$, like in a
stationary Bose condensate. We then create ``cosmological'' 
quasiparticles just by changing 
$g$ (using Feshbach resonances, cf., e.g., Refs.\,\cite{Feshbach}),  
and accordingly change the trap frequencies $\omega_i=\omega$ (in the isotropic case). 
Following Eqs.\,(\ref{deftau}), (\ref{defF}) and 
(\ref{genbEq}), we then have the simple relations  
\begin{equation}
\frac{\omega^2(t)}{\omega_0^2} = \frac{g(t)}{g(0)} =\frac{d\tau_s}{dt}
=\frac1{F(t)} .  \label{EqualSeq}
\end{equation} 
The metric associated with such a thermal 
quasiparticle universe created by ``shaking the trap'' and 
simultaneously changing the interaction appropriately reads, from 
Eq.\,(\ref{blineelement})
\begin{eqnarray}
ds^2 & = &  
\frac{\tilde c \sqrt F }{g(0)} \left[ -{\tilde c}^2 F d\tau_s^2 + 
d{x_i}^2\right].
\end{eqnarray}
Now, defining the scale factor of the BEC quasiparticle universe by the relation   
\begin{equation}
a^2_{\rm scal} \equiv \frac{\tilde c}{g(t)/g(0)},\label{defScaleFactor}
\end{equation}
we have, up to the (irrelevant) factor $1/g(0)$,
\begin{equation}
ds^2 = -a^6_{\rm scal} d\tau_s^ 2 +a^2_{\rm scal} d{x_i}^2.
\end{equation}
This is the form of the metric employed in \cite{BLVFRW}, 
where it was used to calculate cosmological quasiparticle production, 
inspired by a model of Parker \cite{Parker}. 
Note that here a nontrivial scale factor is induced without 
expanding the cloud.
We see from relations (\ref{EqualSeq}) and  (\ref{defScaleFactor})
that the scale factor $a_{\rm scal}$ is in our harmonically trapped case
simply proportional to the ratio of initial and 
instantaneous trapping frequencies, $a_{\rm scal}= 
\sqrt{\tilde c}\, \omega_0/\omega(t)$. 

In Ref.\,\cite{BLVFRW}, a specific choice of the scale 
function $a_{\rm scal}(\tau_s)$ was taken for the 
calculation of the quasiparticle creation process, 
\begin{equation}
a_{\rm scal}^4(\tau_s)=\frac{a_{\rm scal,i}^4+a_{\rm scal,f}^4}2  
+  \frac{a_{\rm scal,f}^4-a_{\rm scal,i}^4}2 
\tanh\left[\frac{\tau_s}{{\tau_s}_0}\right]
\label{eq:prof} 
\end{equation}
where $a_{\rm scal,i}$ and $a_{\rm scal,f}$ are initial and final
scale factors, respectively. 
In the adiabatic approximation, one 
obtains a thermal spectrum \cite{BLVFRW,Parker}, with 
a temperature governed by the inverse laboratory time scale 
$t_0 \propto {\tau_s}_0$ on which 
trapping frequencies, interaction and thus the scale factors change:
\begin{eqnarray}
T_0= \frac1{4\pi t_0} 
\frac{a^4_{\rm scal,i}+a^{4}_{\rm scal,f}} {a^{2}_{\rm scal,f}a^{2}_{\rm scal,i}}.
\label{temperature}
\end{eqnarray}
This temperature is, according to Eq.\,(\ref{defT0}), 
determined by the singular points of the tanh 
function in Eq.\,(\ref{eq:prof}).
The fact that the spectrum is thermal is obtained in \cite{BLVFRW}
for a specific example with a certain form of the 
time dependent interaction. 
We emphasize here that the thermal spectrum is a generic feature 
of adiabatic evolution in harmonically trapped superfluid gases
with temporally varying trapping potential and interactions.

\section{Detection by measuring density-density correlations} 
\label{detect} 
Although the
solutions of the hydrodynamic equations are unique, their {\em interpretation}
in terms of the number of phonons in a given mode 
is subject to all the conceptual
difficulties encountered by the definition of particle states in curved space-times \cite{BirrellDavies}. 
In \cite{fedichev:hawkingPRL}, we have shown that simply by
choosing a specific realization of a quasiparticle (phonon) detector one
can observe thermal quantum ``radiation'' from a de Sitter horizon 
(the Gibbons-Hawking effect \cite{Gibbons}) 
as a purely choice-of-observer related phenomenon, without energy transport 
or dissipation taking place inside the liquid. 
Below, we confine ourselves to the standard (conventional) laboratory
means of particle detection (a CCD camera detecting individual atoms,
rather than phonons), and concentrate on  uniquely defined laboratory frame 
observables, such as the lab frame 
density-density correlations discussed in what follows.

The density fluctuation operator is given by 
\begin{eqnarray}
\delta\tilde{\rho} & = & 
\sum_{n}\sqrt{\frac{1}{2\tilde{V}\epsilon_n \tilde\kappa}}\frac{\partial}{\partial\tau_s}
\left\{
\phi_{n}({\bm x}_{b})
\left[{\hat{a}}_{n}\chi_{n}+{\hat{a}}_{n}^{\dagger}\chi_{n}^{*}\right] 
\right\}\!\!,
\end{eqnarray}
so that the lab-frame density-density correlator $G({\bm x}_{12})
=\langle\delta\tilde{\rho}({\bm x}_{1})
\delta\tilde{\rho}({\bm x}_{2})\rangle/{\cal V}^2$, 
averaged over the initial state is, in the isotropic case, 
\begin{eqnarray}
G({\bm x}_{12}) & = & \sum_{n}\frac{\epsilon_{n}}{2\tilde{V}\tilde{\kappa}}
\frac{\sqrt F}{{\mathcal{V}}^{2}}\phi_{n}\left(\frac{{\bm x}_{1}}b\right)
\phi_{n}\left(\frac{{\bm x}_{2}}b\right)\times 
\\
 & & \hspace*{1em} \times \left[1+2|\beta_{n}|^{2}+2{\rm Re}
\left\{ \alpha_{n}\beta_{n}^{*}e^{-2i\epsilon_{n}\int{d\tau_s}
\sqrt{F}}\right\} \right]. \nonumber 
\end{eqnarray}
 Here the normalization condition (\ref{bognorm}) is used, 
and $r_{12}=|{\bm x}_{12}|=|{\bm x}_{1}-{\bm x}_{2}|\ll|{\bm x}_{1}|,|{\bm x}_{2}|$. 

The TP condition (\ref{TPS}) ensures that the cross-term proportional
to $\alpha\beta^{*}$ averages to zero at large times
$t.$ The term with $1$ in the square brackets describes the evolution
of the vacuum fluctuations and the summation over $n$ is cut off
at the Planckian energy scale: max[$\epsilon_{n}]\sim E_{\rm Pl}=\tilde{\rho} g(0)$.
Accordingly, the corresponding correlation function decays at Planckian distances
$r_{12}/b\sim \tilde c/E_{\rm Pl}$ and is very short-range. Subtracting
the vacuum contribution, we obtain the following expression for the regularized
correlator 
\begin{equation}
G_{{\rm reg}}({\bm x}_{12})=\sum_{n}\frac{\sqrt F \epsilon_{n}}{\tilde V{\mathcal{V}}^{2}\tilde{\kappa}}\phi_{n}\left(\frac{{\bm x}_{1}}{b}\right)\phi_{n}\left(\frac{{\bm x}_{2}}{b}\right)N_{n}.
\end{equation}
 We note that in QFT the regularization procedure does not follow in a unique manner 
from the field theory itself, and can be applied using different assumptions
about the high-energy behaviour of the excitations created from the fundamental ``ether.'' Here, the spectrum (and origin) 
of the TP excitations is well-known, and hence the above regularization
of two-point correlation functions can always be strictly justified. 
This regularization procedure
of course is not limited to density-density correlators only. A similar
technique can be used, for example, to find a regularized energy-momentum tensor.

To be more specific, consider a large Bose-Einstein-condensed gas cloud in
the Thomas-Fermi limit. Then, close to the center of the gas we can
use WKB (plane wave) functions $\phi_{n}$, 
with energies $\epsilon_{k}=\tilde c k$, and 
the regularized Green function is given by 
\begin{equation}
G_{{\rm reg}}(r_{12})=\frac{\tilde{\rho}^{2}}{{\mathcal{V}}^{2}}
\sqrt{\tilde{\rho}g^{3}(0)} \, 
{\mathcal{G}}\left(\frac{T_0 r_{12}}{\tilde c b }\right),
\end{equation}
where the function 
\begin{equation}
{\mathcal{G}} 
=\frac{1}{2\pi^{2}E_{\rm Pl}^{2}}
\int_0^{\infty}\frac{\sin(kr/b)}{r/b}N_{k}k^{2}dk.
\end{equation}
The function ${\mathcal{G}}$
reaches its maximum for $r_{12}=0$, so that the signal to noise ratio 
is maximally 
\begin{equation}
\frac{G_{{\rm reg}}(0)}{\rho^{2}}\sim \sqrt{\tilde{\rho}a_s^{3}(0)}
\left(\frac{T_{0}}{E_{\rm Pl}}\right)^{4}. \label{relCorrel}
\end{equation}
The above discussion 
shows that even from a small ``noise'' signal,  
one can extract the relevant features of the quantum state 
of the gas cloud (for a more detailed discussion cf. \cite{Altman}). 
In order to be measurable, the quantity (\ref{relCorrel}) has to be
of the order of a few percent. This can in principle be achieved by using initially
dense clouds with strong interparticle interactions.  
Finally, we mention that a similar, i.e.,
velocity-velocity instead of density-density ``noise'' 
correlation function has already been measured in the 
experiments of \cite{Hellweg}. 

\section{Conclusion} 
In the present investigation, we have derived the general scaling 
equations for harmonically trapped 
superfluid Bose and Fermi gases, and related these, in particular,  
to quasiparticle metric tensors of the de Sitter and 
Friedmann-Robertson-Walker type, familiar from a cosmological context. 
The quasiparticle creation in a harmonically
trapped superfluid gas, by changing 
interaction {and} trapping simultaneously in an appropriate manner, 
can therefore be described in a general framework, and 
be interpreted to be analogous to the particle creation occuring during 
rapid expansion of the cosmos. 
In particular, it was found that for 
a readily experimentally available case, the harmonically trapped, 
dilute superfluid Bose gas, a FRW type metric can be induced if trapping
frequency $\omega$ and interaction coupling $g$ are changed such
that $\omega^2(t) \propto g(t)$, without expanding the gas. 
The cosmological scale factor in this case 
is inversely proportional to the trap frequency, $a_{\rm scal} (t) \propto 1/\omega(t)$.

If the frequency mixing leading to quasiparticle
creation can be described 
in the WKB approximation, generally a thermal 
distribution is found, where the temperature is determined by the 
singular points of the scaling factors given by Eq.\,(\ref{defF}),  
in the complex plane of scaling time $\tau_s$.

We finally stress that, in contrast to a
typical cosmological calculation, hydrodynamic fluctuations 
in a laboratory experiment always have a well-defined initial 
state in the lab frame, with time coordinate $t$. 
Therefore, ambiguities of the final quantum state as regards the
dependence of its particle content on 
the initial conditions imposed on the ``vacuum'' can be ruled out: 
There exists the preferred lab frame vacuum, uniquely prescribing 
the initial particle content of the quantum field.

\begin{acknowledgments}
P.\,O.\,F. has been supported by the Austrian Science Foundation
FWF and the Russian Foundation for Basic Research, and U.\,R.\,F.
by the FWF. They both were supported by the 
ESF Programme {}``Cosmology in the Laboratory,'' and gratefully acknowledge
the hospitality extended to them during the Bilbao workshop.  
We thank J.\,I. Cirac, U. Leonhardt, R. Parentani, 
R. Sch\"{u}tzhold, M. Visser, G.\,E. Volovik, and P. Zoller 
for helpful correspondence and discussions. 
\end{acknowledgments}

\end{document}